\documentclass[nonacm=true,acmtog=false,anonymous=false,review=false]{acmart}
\acmSubmissionID{666}

\pdfoutput=1

\usepackage{booktabs}
\usepackage{colortbl}
\usepackage{graphicx}
\usepackage{subcaption}
\usepackage{changepage}
\usepackage{wrapfig}
\usepackage{enumitem}
\usepackage{mathtools} 

\usepackage{algorithm}
\usepackage[noend]{algpseudocode}

\algnewcommand{\LineComment}[1]{\Statex $\quad\;\;$\(\triangleright\) #1}

\definecolor{WHITE}{gray}{1.0}
\definecolor{GRAY}{gray}{0.9}

\newcommand{\ie}{\textit{i.e.,}~}

\pdfpageattr {/Group << /S /Transparency /I true /CS /DeviceRGB>>}

\let\orgautoref\autoref

\def\secnospace~{\S{}}
\renewcommand{\autoref}
        {\def\equationautorefname{Eq.}%
         \def\figureautorefname{Fig.}%
         \def\subfigureautorefname{Fig.}%
         \def\algorithmautorefname{Alg.\@}%
         \def\Itemautorefname{Item}%
         \def\tableautorefname{Table}%
         \def\sectionautorefname{\secnospace}%
         \def\subsectionautorefname{\secnospace}%
         \def\subsubsectionautorefname{\secnospace}%
         \def\chapterautorefname{\secnospace}%
         \def\partautorefname{Part}%
         \orgautoref}

\def\BibTeX{{\rm B\kern-.05em{\sc i\kern-.025em b}\kern-.08em
    T\kern-.1667em\lower.7ex\hbox{E}\kern-.125emX}}

\newcommand{\figTeaser}{
\begin{teaserfigure}
  	\captionsetup[subfigure]{}
    \centering
    \subcaptionbox{\label{fig:teaserBunny}} {
        \includegraphics[height=1in,trim={0in 0in 0in 0in},clip]{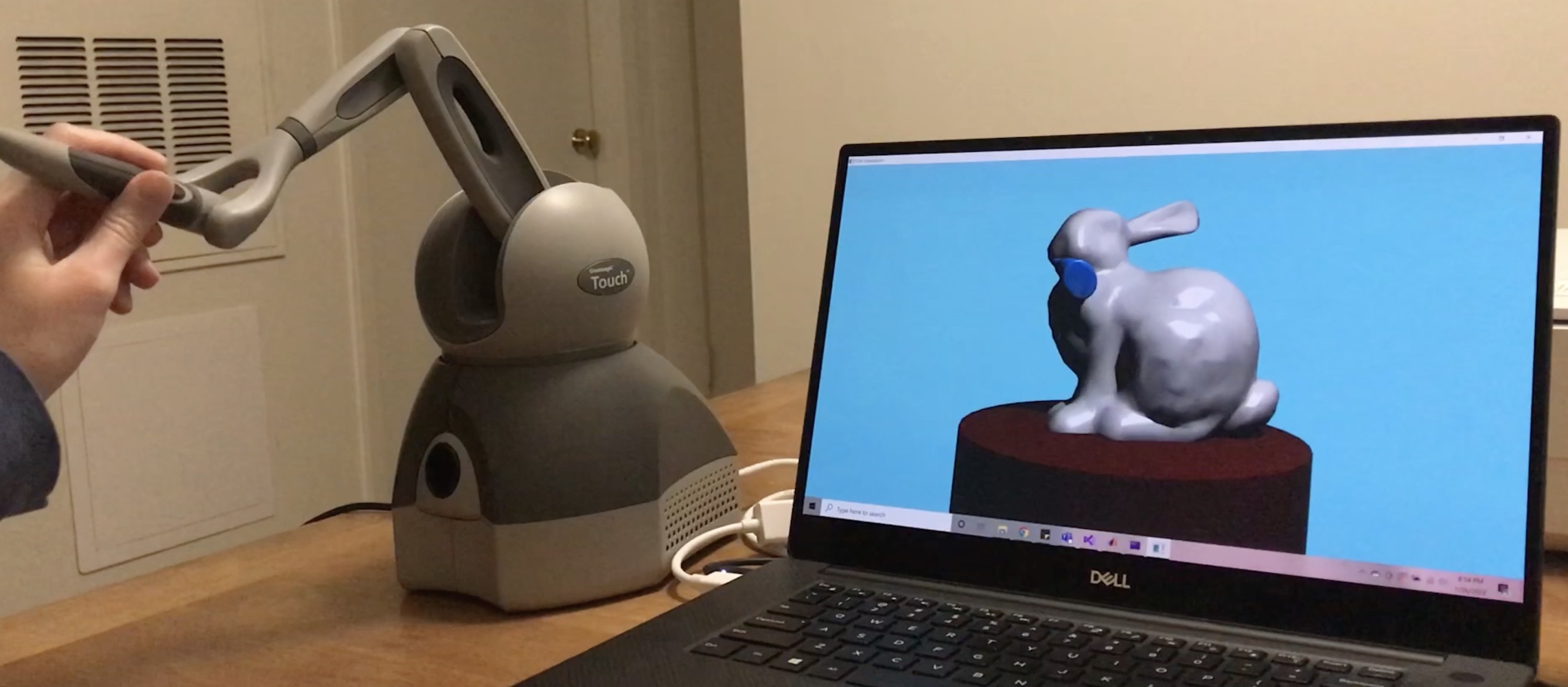}
    }
    \subcaptionbox{\label{fig:teaserCloth1}} {
        \includegraphics[height=1in,trim={2.5in 0in 2.5in 0in},clip]{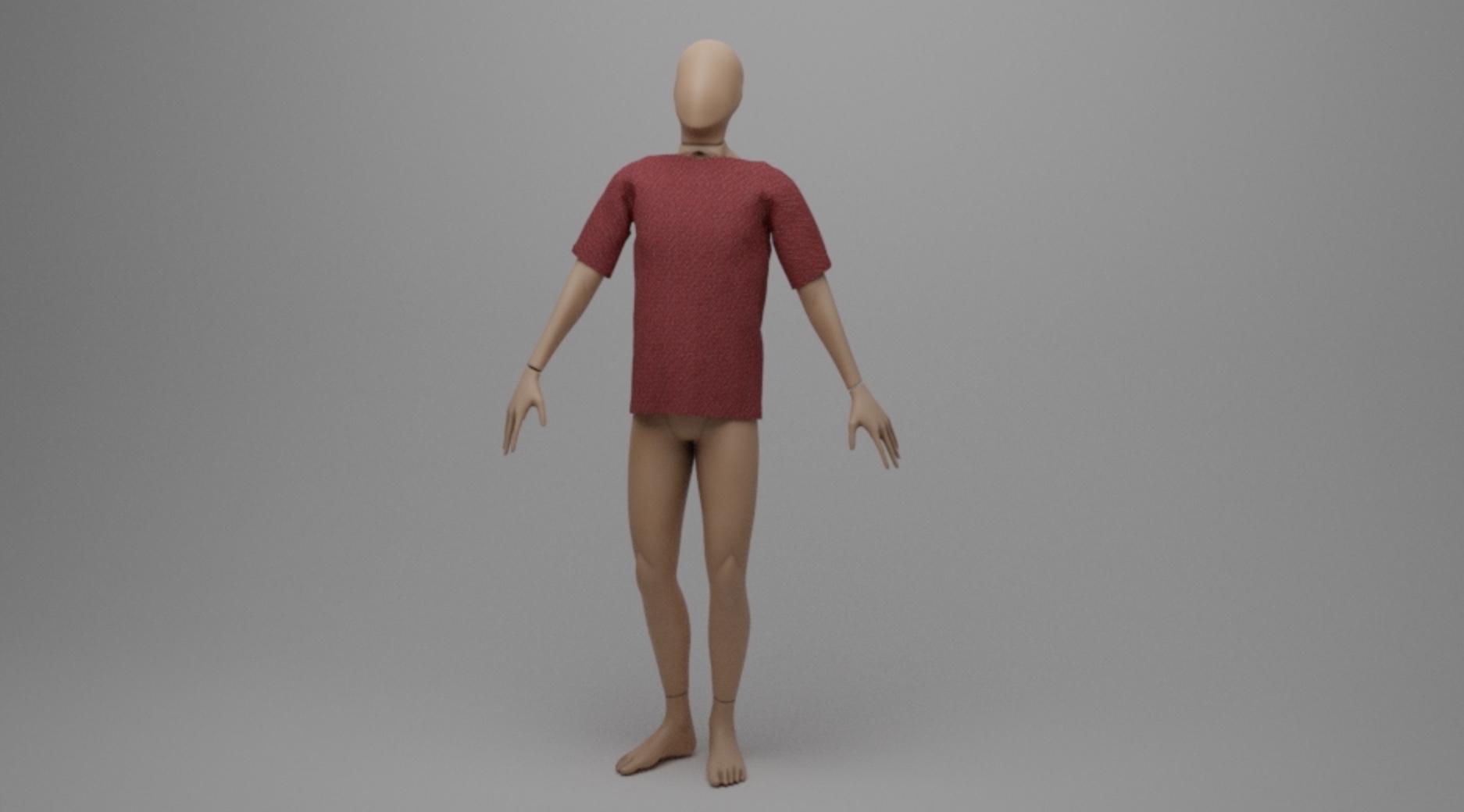}
    }
    \subcaptionbox{\label{fig:teaserCloth2}} {
        \includegraphics[height=1in,trim={2.5in 0in 2.5in 0in},clip]{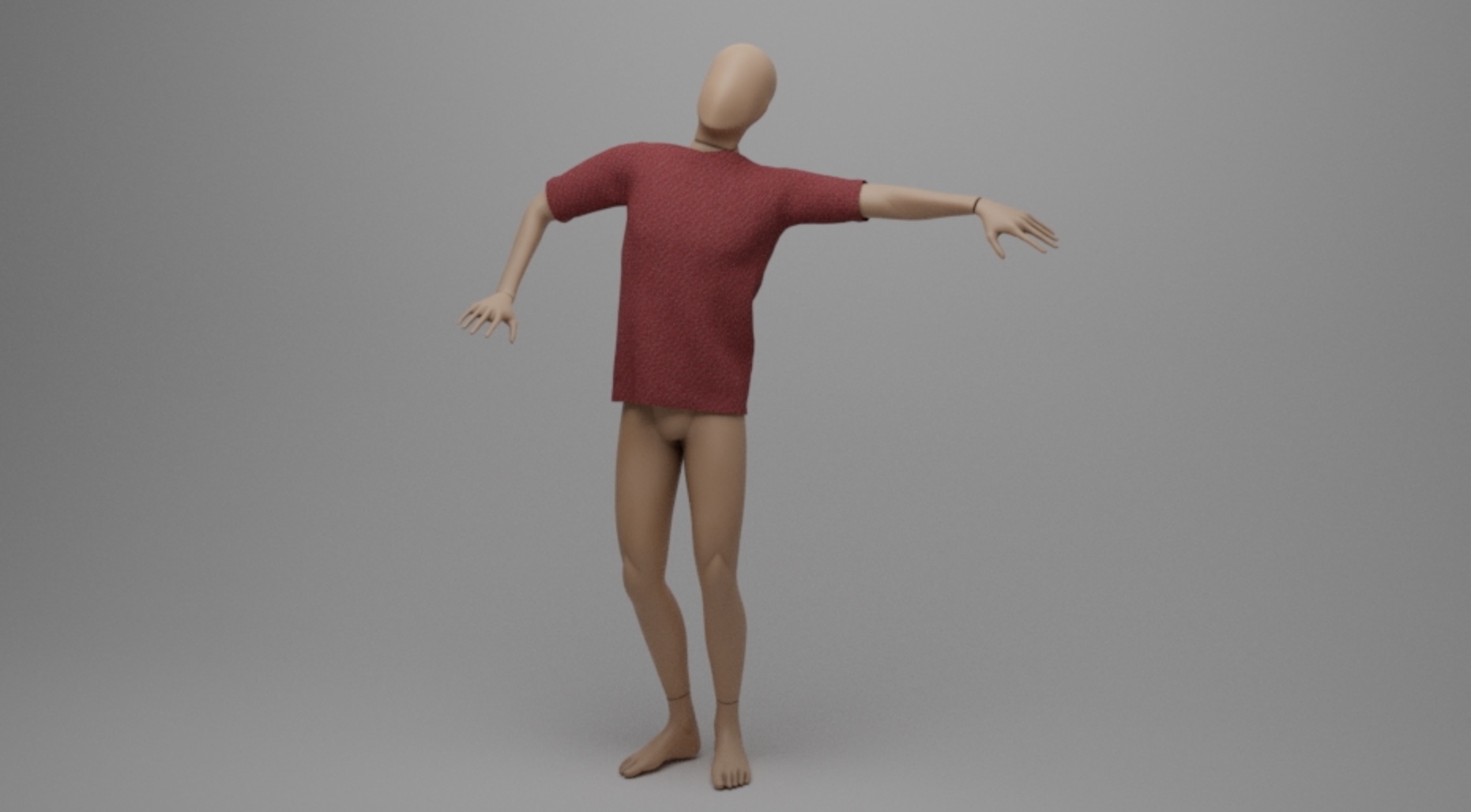}
    }
    \subcaptionbox{\label{fig:teaserCloth3}} {
        \includegraphics[height=1in,trim={2.5in 0in 2.5in 0in},clip]{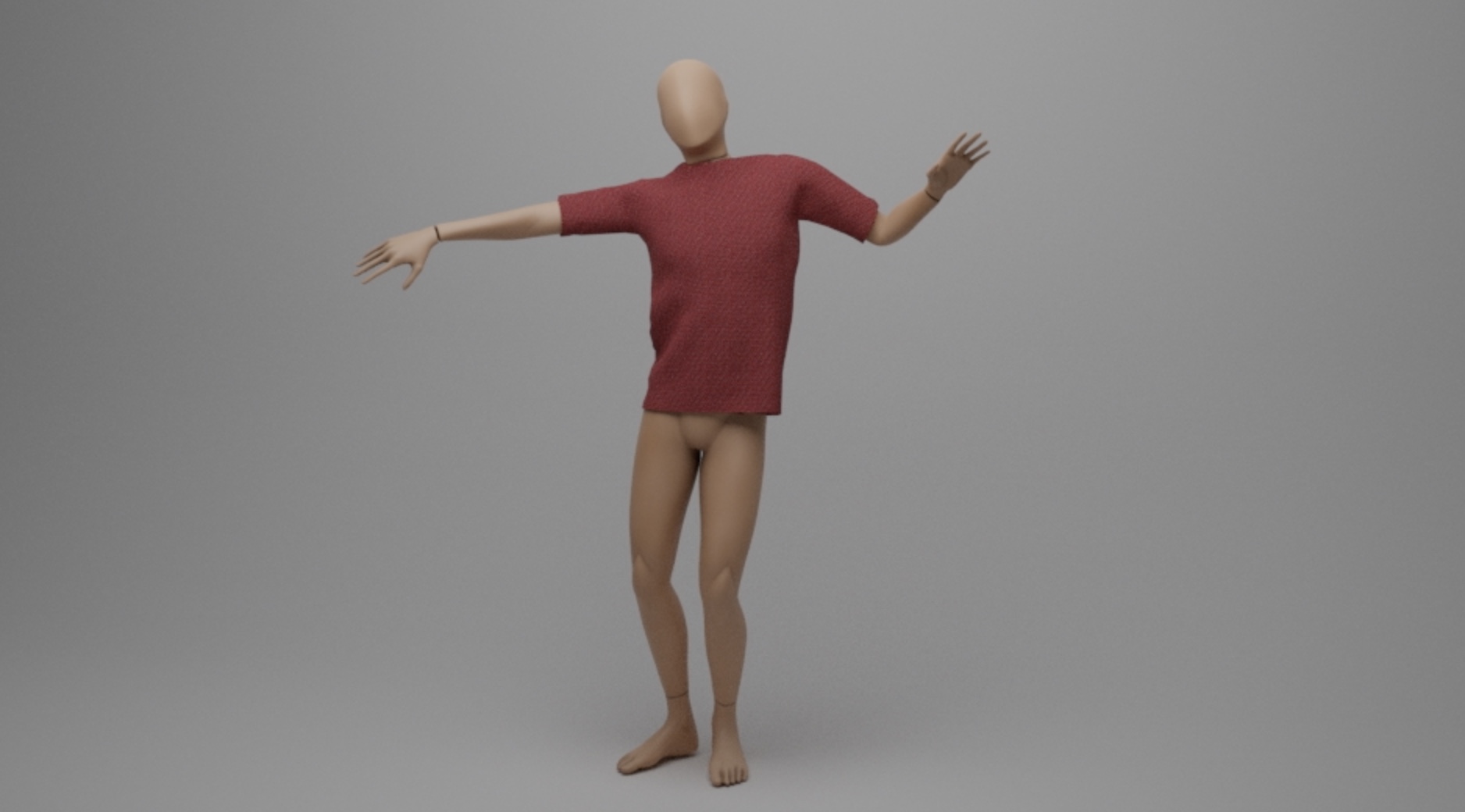}
    }
    \caption{
        Demonstrations of our neural collision detector for deformable objects.
        (a) Haptics with a finite element mesh.
        (b-d) Cloth on a skinned character.
    }
    \label{fig:teaser}
\end{teaserfigure}
}

\newcommand{\figBunnyLRQvsT}{
  \begin{figure}[tb]
    \centering
    \includegraphics[width=0.99\columnwidth,trim={1.25in 3in 1.65in 3.15in},clip]{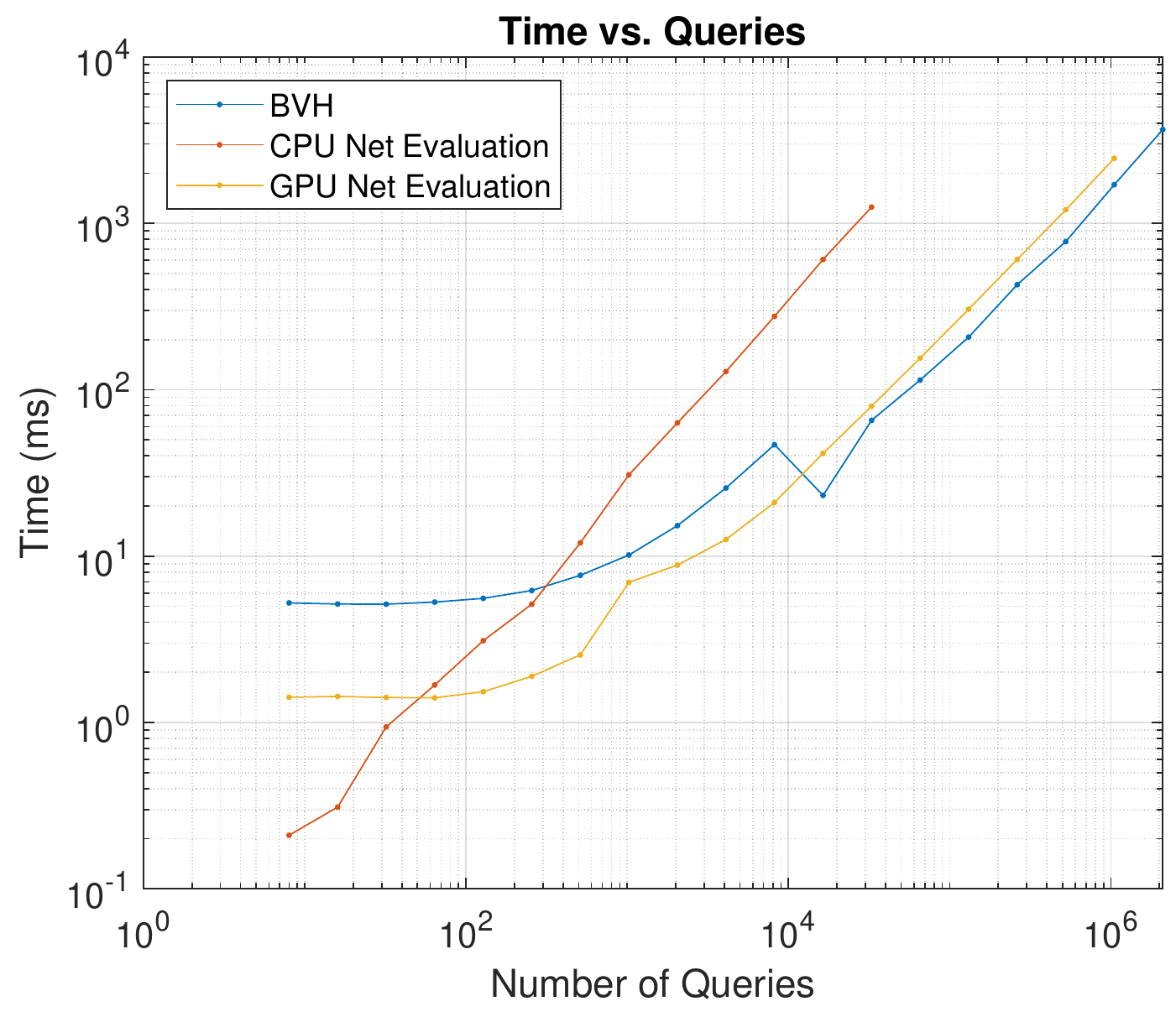}
    \caption{
      Time to query the signed distance of a point, either by evaluating the trained bunny network or using a BVH.
    }
    \label{fig:FRBunnyQvsT}
  \end{figure}
}

\newcommand{\figPersonQvsT}{
  \begin{figure}[tb]
    \centering
    \includegraphics[width=0.99\columnwidth,trim={1.25in 3in 1.65in 3.15in},clip]{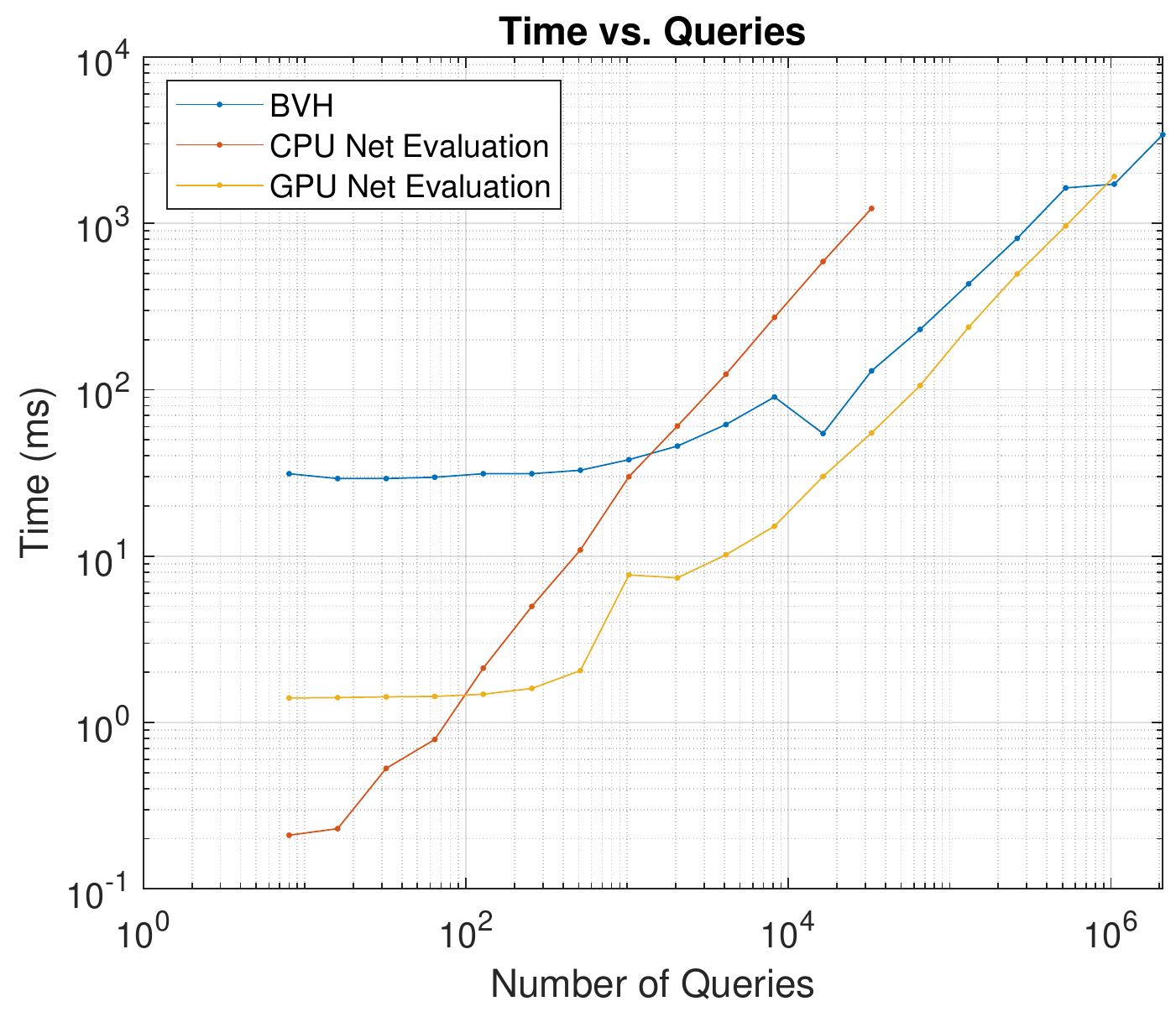}
    \caption{
      Time to query the signed distance of a point, either by evaluating the trained character network or using a BVH.
    }
    \label{fig:PersonQvsT}
  \end{figure}
}

\newcommand{\figBunnyLevelsSmallNet}{
  \begin{figure}[tb]
    \centering
    \includegraphics[width=0.32\columnwidth,trim={0.6in 0.2in 0.2in 0.2in},clip]{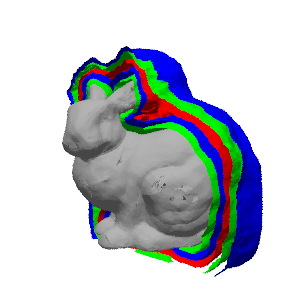}
    \includegraphics[width=0.32\columnwidth,trim={0.6in 0.2in 0.2in 0.2in},clip]{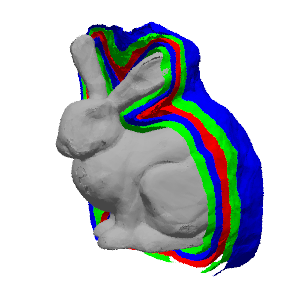}
    \includegraphics[width=0.32\columnwidth,trim={0.6in 0.2in 0.2in 0.2in},clip]{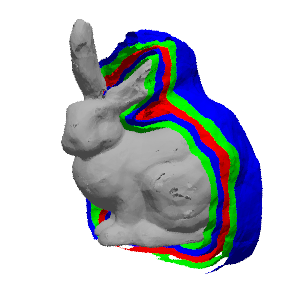}
    \includegraphics[width=0.32\columnwidth,trim={0.6in 0.2in 0.2in 0.2in},clip]{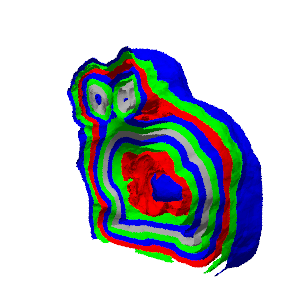}
    \includegraphics[width=0.32\columnwidth,trim={0.6in 0.2in 0.2in 0.2in},clip]{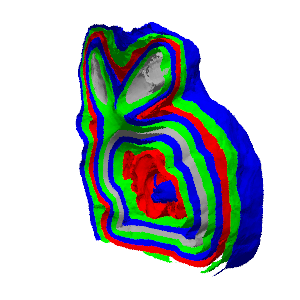}
    \includegraphics[width=0.32\columnwidth,trim={0.6in 0.2in 0.2in 0.2in},clip]{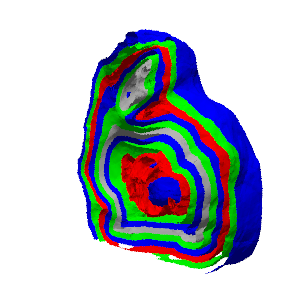}
    \caption{
        Level sets of a learned bunny at three deformations, with and without the full zero level set rendered. Level sets are in increments of $0.05$.
    }
    \label{fig:BunnyLevels}
  \end{figure}
}

\newcommand{\figPersonLevelsSmallNet}{
  \begin{figure}[tb]
    \centering
    \includegraphics[width=0.32\columnwidth,trim={0.6in 0.6in 0.3in 0.2in},clip]{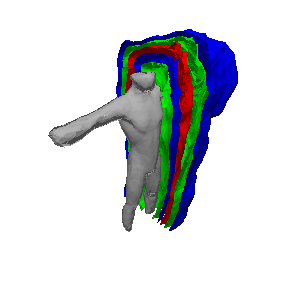}
    \includegraphics[width=0.32\columnwidth,trim={0.6in 0.6in 0.3in 0.2in},clip]{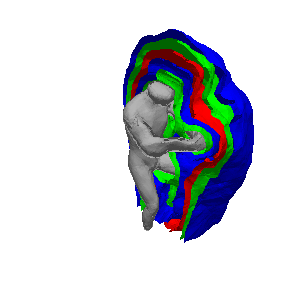}
    \includegraphics[width=0.32\columnwidth,trim={0.6in 0.6in 0.3in 0.2in},clip]{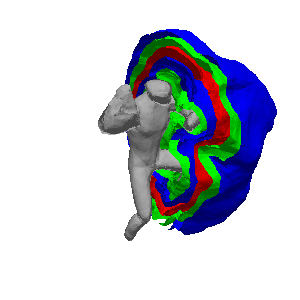}
    \includegraphics[width=0.32\columnwidth,trim={0.6in 0.6in 0.3in 0.2in},clip]{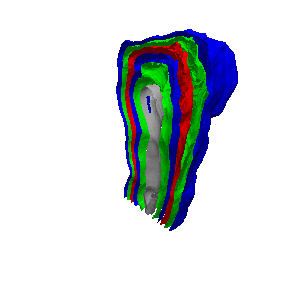}
    \includegraphics[width=0.32\columnwidth,trim={0.6in 0.6in 0.3in 0.2in},clip]{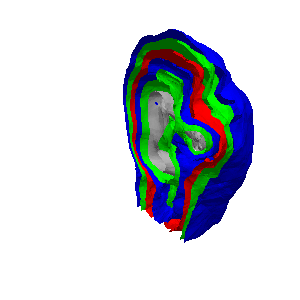}
    \includegraphics[width=0.32\columnwidth,trim={0.6in 0.6in 0.3in 0.2in},clip]{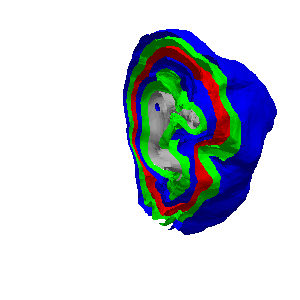}
    \caption{
        Level sets of a learned person at three deformations, with and without the full zero level set rendered. Level sets are in increments of $0.075$.
    }
    \label{fig:PersonLevels}
  \end{figure}
}

\newcommand{\figBunnyDrop}{
\begin{figure}[tb]
  \centering
    \includegraphics[width=0.47\columnwidth,height=1in,trim={0in 0in 0in 0in},clip]{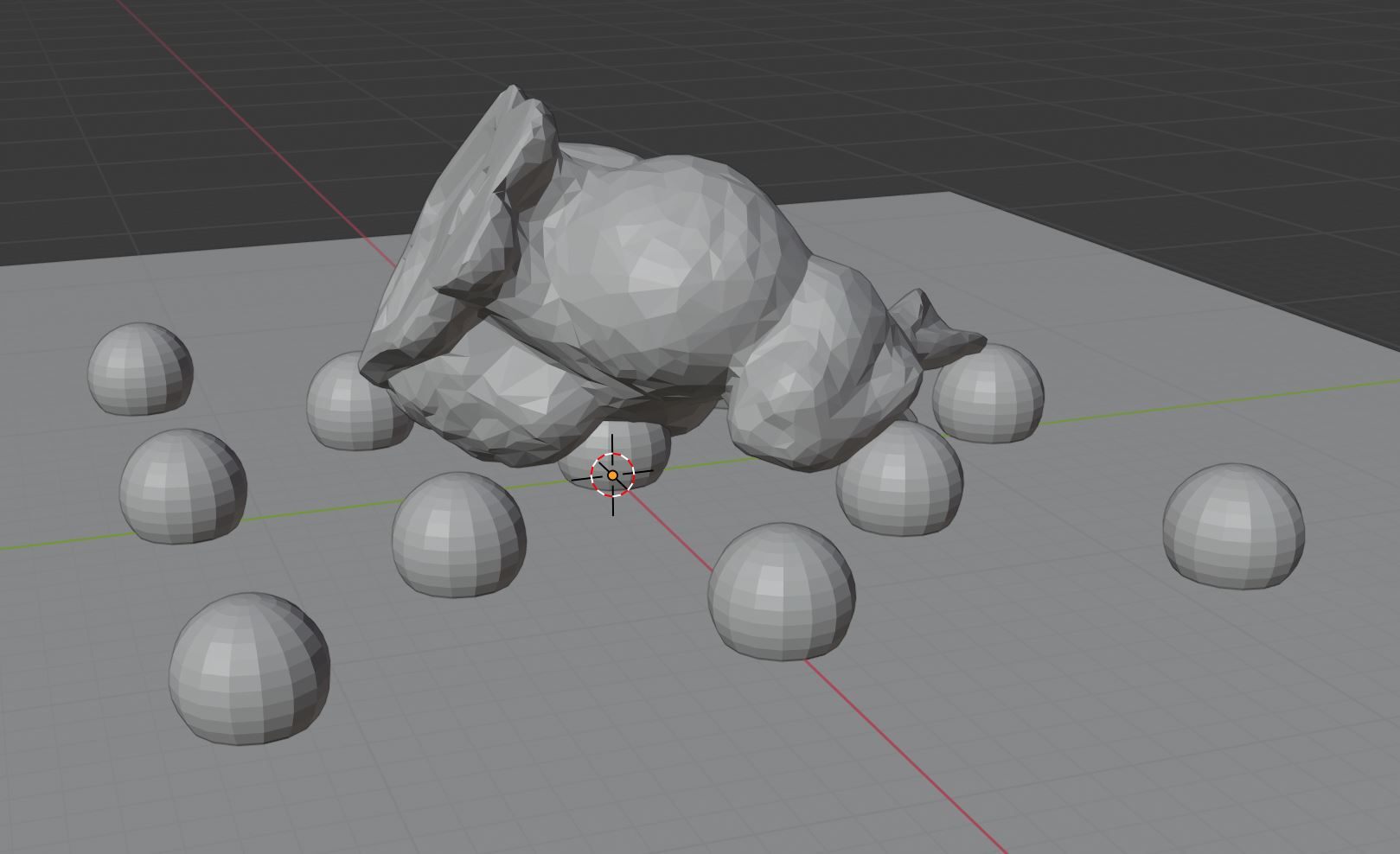}
    \includegraphics[width=0.47\columnwidth,height=1in,trim={0in 0in 0in 0in},clip]{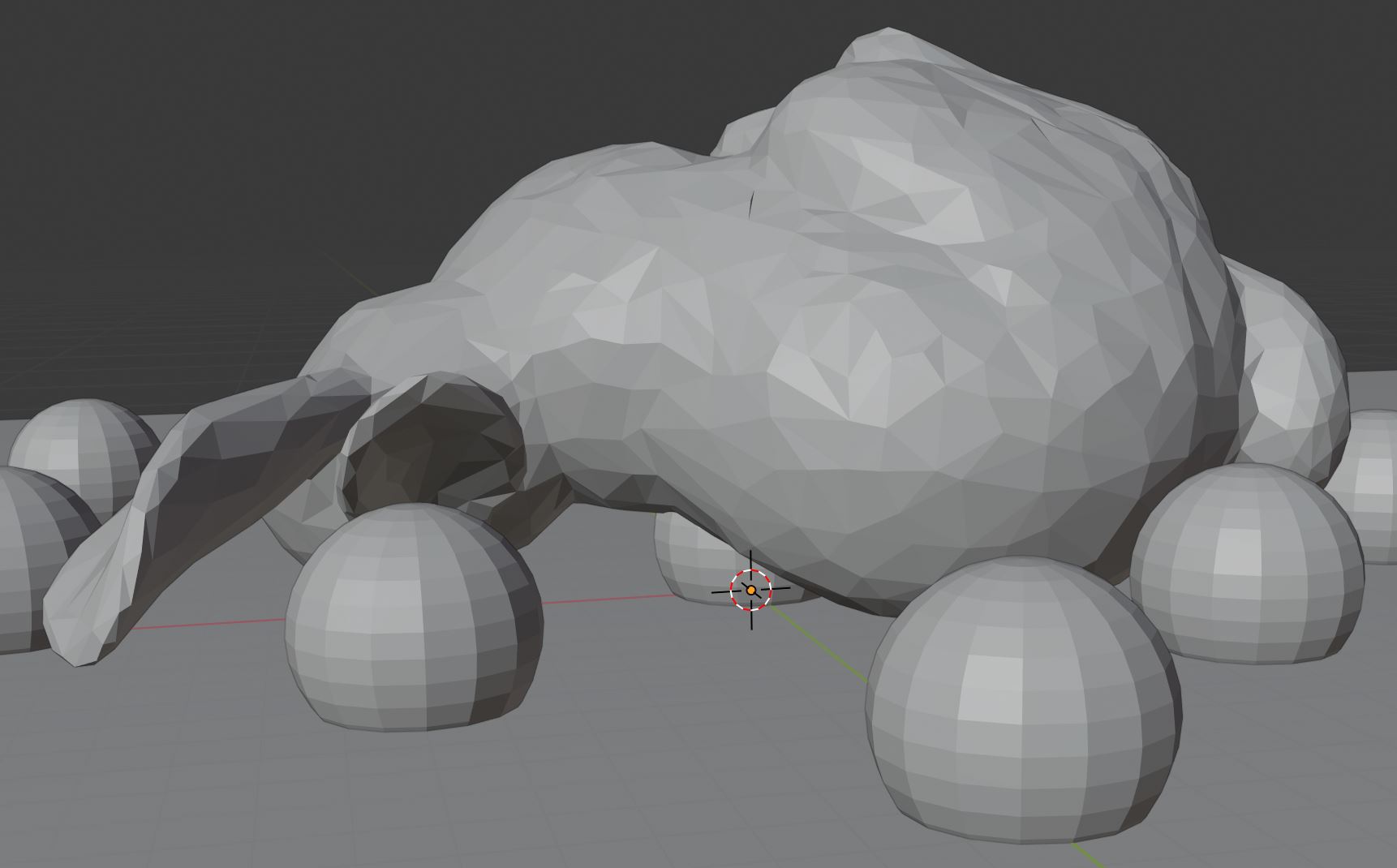}
    \caption{
      Example training poses for the volumetric bunny object.
    }
    \label{fig:BunnyDrop}
  \end{figure}
}

\setcitestyle{nosort,square} 

\acmJournal{TOG}
\acmVolume{1}
\acmNumber{1}
\acmArticle{1}
\acmYear{2022}
\acmMonth{1}

\begin{document}
\title{Neural Collision Detection for Deformable Objects}


\author{Ryan S. Zesch}
\affiliation{%
 \institution{Texas A\&M University}
 \country{USA}}
\email{rzesch@tamu.edu}

\author{Bethany R. Witemeyer}
\affiliation{%
 \institution{Texas A\&M University}
 \country{USA}
}
\email{bethanywitemeyer@tamu.edu}

\author{Ziyan Xiong}
\affiliation{%
 \institution{Texas A\&M University}
 \country{USA}
}
\email{zyxiong@tamu.edu}

\author{David I.W. Levin}
\affiliation{%
 \institution{University of Toronto}
 \country{Canada}
}
\email{diwlevin@cs.toronto.edu}

\author{Shinjiro Sueda}
\orcid{0000-0003-4656-498X}
\affiliation{%
 \institution{Texas A\&M University}
 \country{USA}}
\email{sueda@tamu.edu}

\renewcommand\shortauthors{Zesch, R. et al.}

\begin{abstract}
We propose a neural network-based approach for collision detection with deformable objects.
Unlike previous approaches based on bounding volume hierarchies, our neural approach does not require an update of the spatial data structure when the object deforms.
Our network is trained on the reduced degrees of freedom of the object, so that we can use the same network to query for collisions even when the object deforms.
Our approach is simple to use and implement, and it can readily be employed on the GPU.
We demonstrate our approach with two concrete examples: a haptics application with a finite element mesh, and cloth simulation with a skinned character.
\end{abstract}

%
%

%
%


\figTeaser{}

\maketitle

\section{Introduction}
\label{sec:intro}


Collision detection is an important component of the computer animation pipeline, but an efficient approach for deformable objects remains a challenge \cite{Teschner2005}.
Two of the most commonly used approaches for collision detection are Signed Distance Fields (SDF) and Bounding Volume Hierarchy (BVH).
An SDF is highly efficient for a rigid object, but it cannot be easily used for complex deformations.
On the other hand, a BVH can be used for arbitrarily deforming meshes, but it must maintain a spatial data structure that must be updated whenever the object deforms.
In this work, we propose an approach that, like a BVH, can be used for complex deformations but, like an SDF, does not require an update of the underlying spatial data structure when the object deforms.

Our approach is to build a neural network that is trained on a ``code'' of the deformed mesh.
This code can be any reduced representation of the deformation.
In this work, we show two examples: linear modes for a finite element mesh and joint angles for a skinned mesh.
To generate the training data, we use as input a set of deformed full-space meshes along with their reduced codes.
For each mesh, we sample the distance field around the mesh and store the computed distances as the output corresponding to the input query points and the code.
This allows us to train a network that can be used to compute the signed distance for a deforming object.
At runtime, we evaluate the trained network using the code that corresponds to the current deformed mesh, along with the query point.

Our approach works well when the number of distance queries per deformed mesh is relatively small.
Since the initial cost of updating the spatial data structure at each time step is amortized over the queries, as the number of queries increases, there is a point at which BVH outperforms our approach.
However, since the network evaluation code involves only matrix multiplications and activation functions, our approach can be readily implemented on a wide range of hardware, including the GPU.
On the other hand, it is non-trivial to implement a BVH on the GPU.
Due to performance penalties induced by global memory accesses for scattered queries, GPU BVHs require stackless implementations to achieve peak performance~\citep{Binder2016}. 
Our neural method requires no additional implementation to attain excellent speed-ups on the GPU as it trivially leverages development tools optmized for neural network execution.
Furthermore, updating an existing BVH is also non-trivial on the GPU, and getting a tight bound on a general deforming model remains challenging.
Specialized implementations exist for static scenes and ray tracing applications~\citep{karras2013fast}. 
These limitations exist even for new GPU BVH construction hardware~\citep{Viitanen2018}.
Our neural approach to collision detection has none of these limitations and can be readily deployed for collision detection in deformable physics simulations.

\section{Related Work}
\label{sec:related}

Collision detection has been an active area of research in many fields including graphics, robotics, and vision, with several survey papers spanning multiple decades \cite{Lin1998,Teschner2005,Haddadin2017}.
We refer the reader to these excellent surveys for an overview of various techniques.
One of the most popular approaches for deformable objects collisions is BVH.
If the modes of deformation are known \textit{a priori}, then the BVH can be updated very efficiently \cite{James2004}. 
However, as mentioned in the introduction, updating a tightly fitting BVH for a generally deforming object remains difficult, especially on the GPU.
Image-based methods work well with deformable objects and naturally run on the GPU \cite{Faure2008,Wang2012}, but these methods cannot be readily be incorporated into other simulation frameworks.
Another approach is to deform an SDF based on the object's mesh \cite{Fisher2001,McAdams2011,Macklin2020}.
However, with these methods, a BVH is still required to find the region or the cell that contains the query point.
Deformed SDFs have also been used for deformed sphere tracing and simple collision detection \cite{Seyb2019}, but such methods have limited applicability to general collision detection because they cannot evaluate the underlying implicit surface at an arbitrary point in deformed space.

Recently, approaches based on neural fields have become extremely popular \cite{Xie2021}.
Of these, implicit shape representation through occupancy or signed distance fields are highly relevant to collision detection.
\citet{park2019deepsdf} showed that, with their Coded Shape DeepSDF approach, they can build a highly effective implicit representation of non-rigid 3D geometry.
Concurrent work by \citet{Mescheder2019} and \citet{Chen2019} used neural networks for occupancy fields. All of these works use neural approaches for various visual applications, such as shape completion, interpolation, and 3D reconstruction.
We use a similar idea but for collision detection for deformable objects for animation.

\section{Methods}
\label{sec:methods}

In this section, we describe the training and runtime details of our two concrete applications: (\autoref{sec:methods_haptics}) deformable haptics and (\autoref{sec:methods_cloth}) skinned cloth.
We will first give the full description of the haptics application, and then refer to some of the shared details when we give the description of the cloth application.

\subsection{Haptics with FEM}
\label{sec:methods_haptics}

In our haptics application, we use the finite element method (FEM) to simulate a deformable volumetric object.
We interact with this object with another, lower-resolution volumetric object that is driven by a haptics device.
\autoref{fig:teaserBunny} shows our setup involving a bunny and a blue hand (under the head of the bunny).
Even though both objects are soft, volumetric FEM objects, we will call the bunny the ``volumetric object'' and the hand the ``haptics object.''
For our application, we perform a one-way collision query---we check if the vertices of the haptics object are colliding with the volumetric object.
We also assume that both objects are relatively smooth, and so do not check for edge-edge collisions.

To perform collision detection, the network must use the information about the current deformation of the volumetric object.
However, using the full-space information (\ie the positions of all the vertices) of the deformed mesh as the input to the neural network quickly becomes untractable as the resolution of the mesh increases.
We therefore project the full-space information down to a reduced space, and use the resulting reduced coordinates as the input ``code'' to the network.
In our current implementation, we use linear modes \cite{Pentland1989}, but any reduced coordinates can be used, such as nonlinear modes \cite{Sifakis2012} or nested cages \cite{Sacht2015}.
It is important to note, however, that \textit{this reduction is used only for the code for the collision network}---our finite element simulator runs in the full space of deformations.
Our neural collision approach can be used regardless of what space the rest of the simulation uses.

\subsubsection{Training}


Let $x$ denote the current, deformed vertex positions of the volumetric object, and $X$ the undeformed vertex positions at rest.
As a preprocessing step, we compute the linear modal basis vectors $U$ for $m$ modes of the object at its rest pose $X$ \cite{Sifakis2012}.
We then generate a set of $n$ training poses $x_i$ of the object undergoing a range of deformations \textit{using the full-space simulator}.
These training poses are generated by dropping the volumetric object onto the ground with spheres of various sizes protruding out, as shown in \autoref{fig:BunnyDrop}.
We go through this set of poses, and compute $\bar{x}_i$, the vertex positions with the rigid modes removed, by rigidly aligning $x_i$ to $X$ \cite{Sifakis2012}.
We then compute the corresponding codes $z_i$ as
\begin{equation}
\label{eq:FEMcode}
    z_i = U^\top (\bar{x}_i - X),
\end{equation}
which gives us a set $(\bar{x}_i, z_i)$ for $i = 1, 2, \cdots, n$ of rigidly-aligned full-space poses and their corresponding reduced-space codes.
We then construct an SDF around each $\bar{x}_i$ \cite{Batty2015}.
Finally, we query the SDF at various locations, using the following heuristics:
\begin{itemize}
    \item Surface samples
    \item Nearby samples, within a tight normal distribution of the surface
    \item Uniform grid samples, within the bounding box containing all $\bar{x}_i$.
\end{itemize}
These sample types are used in a $2:2:1$ ratio.
This gives us a set of training samples $(q_k, z_k) \rightarrow{} d_k$ for $k = 1, 2, \cdots, |q|$, where $q$ is the query point, $z$ is the code, $d$ is the signed distance, and $|q|$ is the total number of queries across all poses.
Before training, all query points $q_k$ are uniformly scaled and translated to lie within $[-1,1]^3$, and codes $z_k$ are normalized element-wise to $[-1,1]$.

\figBunnyDrop{}

The network architecture used is a multilayer perceptron (MLP) with ReLU non-linearities and a $\tanh$ final activation function.
We use the clamped $L_1$ loss between the network output $f(q,z)$ and the actual distance $d$, as suggested by DeepSDF \cite{park2019deepsdf}:
\begin{equation}
    \mathcal{L}(f(q,z),d) = \| \textnormal{clamp}(f(q,z),\pm\delta) - \textnormal{clamp}(d,\pm\delta) \|_1.
\end{equation}
For all of our results, we use $\delta=0.1$. 

\subsubsection{Runtime}

At runtime, the haptics simulator calls the collision detector with a set of query points $q$ from the haptics object and the current deformed positions $x$ of the volumetric object.
We first construct the code $z$ using \autoref{eq:FEMcode}.
Then we evaluate the network using the query point $q$ and the code $z$ to get the corresponding collision distance $d$.
To increase throughput, we batch all queries together by forming an input matrix with the code repeated for each query point:
\begin{equation}
\label{eq:batch}
    \begin{pmatrix}
    q_1 & q_2 & q_3 & \hdots\\
    z & z & z & \hdots
    \end{pmatrix}.
\end{equation}
The returned vector $(d_1 \; d_2 \; d_3 \; \hdots )$ then contains the corresponding signed distances.
To obtain the collision normal, we perform a backward pass of the network to compute the derivative of the distance with respect to the query point.

In addition to the collision normal and depth, the haptics simulator requires the ID of the triangle of the volumetric object to apply the force to.
To obtain this information, we perform a linear scan of the surface triangles.
For each triangle, we check if the ray formed by the query point and the collision normal intersects the triangle.
Fortunately, this procedure parallelizes well---we observed near linear scaling with the number of physical cores.

Due to the use of reduced coordinates, there is potentially a projection error between the reduced codes $z$ and the full-space mesh positions $x$.
This means that there are multiple $x$ that could map to the same $z$.
During the construction of the SDF, we may generate two data points $(\bar{x}_i, z_i)$ and $(\bar{x}_j, z_j)$ with $\bar{x}_i \ne \bar{x}_j$ but with $z_i = z_j$.
In such cases, the network would not be able to reproduce these individual data points---instead, the network would produce a smoothed result between them.
With our haptics application, we did not notice this problem, but depending on the application, this problem may become pronounced.
Increasing the size of the reduced space or using a better set of degrees of freedom may be needed to resolve this issue.

\subsection{Cloth with Skinning}
\label{sec:methods_cloth}

For our next application, we use our neural collision detector to compute the collisions between dynamic cloth and a skinned character.
In our current implementation, we use linear blend skinning \cite{Magnenat-Thalmann1988}, but any skinning method can be used without any modifications to the rest of the pipeline.
We again perform one-way collision detection---we check if the vertices of the cloth are colliding with the skinned character.
We also assume that the skinned character is smooth, and so do not check for edge-edge collisions.

Our target use case is not high-fidelity cloth simulation but rather virtual try-on or games.
We assume that we already have a set of animations (\ie sequence of joint angles) that will be applied to the character, though this is not a hard requirement.

\subsubsection{Training}

With skinning, the most natural code for the network is the set of joint angles.
Unlike with our haptics application, there is no projection error with this code---if we know the code, we know the shape of the skinned character without ambiguity.

Assuming that the cloth does not interact with the feet, hands, and head, we do not include their joint angles in the code.
Omitting the hands is particularly helpful, since the hands often have many degrees of freedom.
We further reduce the size of the code by dropping any degrees of freedom that remain constant throughout all of the animation sequences.

In order to generate training data, the character generates a set of $n$ training poses $x_i$ from animation data.
These training poses are expressed in the local space of the root joint, so that rigid motion are automatically accounted for.
We construct and sample SDFs for each of the poses as described in \autoref{sec:methods_haptics}.
Before training, all query points $q_i$ are uniformly scaled and translated to lie within $[-1,1]^3$, and codes $z_i$ are scaled by $\frac{1}{\pi}$ so they lie within $[-1,1]$.

\subsubsection{Runtime}

At runtime, query points are transformed from world coordinates to the model's root's frame of reference. Then, the cloth simulator calls the collision detector with the set of cloth vertices $q$ and the current joint angles $z$.
We again batch the queries before passing them to the network for evaluation to increase throughput.
The network returns distances $d$ and the corresponding collision normals, which are then used to apply forces to drive the cloth. 
Unlike with our haptics application, we do not require the ID of the colliding triangle, since we do not need to apply any collision forces to the skinned character.

\section{Results}
\label{sec:results}

The networks were trained in PyTorch \cite{PyTorch2019} on a Titan RTX GPU with 24 GB of RAM. 
We use the Adam optimizer with the learning rate of $1\times 10^{-4}$ \cite{Kingma14}.
For each application, we use an MLP with 8 layers and 128 neurons per layer.
For the synthetic benchmark, we used a consumer laptop with an Intel\textsuperscript{\textregistered} Core\textsuperscript{\texttrademark} i7-7700 CPU @ 3.60 GHz with 16 GB of RAM.

\subsection{Haptics with FEM}
\label{sec:results_bunny}

The bunny mesh has 1,883 nodes, 3,122 surface triangles, and 6,726 tetrahedra; the hand mesh has 582 nodes, 1,000 surface triangles, and 2,018 tetrahedra.
For training, we use 1,000 poses with 10,000 SDF samples per pose.
We use 128 modes as the reduced code for the network, which we found to be sufficient for our haptics application.
However, more modes may be required to cover highly local deformations that may be needed for some other applications.

Interestingly, we are able to change the stiffness/mass parameters in the online FEM simulation without having to retrain the network, even though the network was trained with the linear modes constructed from a specific set of mass and stiffness values.
This allows us to use a set of simulation settings that are well suited for training and another set for runtime.
During training, we used the fully implicit Euler integrator \cite{Hairer2006} with high density, high Poisson's ratio (0.49), medium Young's modulus, and high damping to ensure sufficient deformation while keeping the simulation stable.
Then for runtime, we used the linearly implicit Euler integrator \cite{Baraff1998} with medium density (1/6th), lower Poisson's ratio (0.40), lower Young's modulus ($\sim$1/10th), and lower damping ($\sim$1/10th) to better enable compelling real-time interactions.

\figBunnyLevelsSmallNet{}

The visualizations of the levelsets of the learned SDF are shown in \autoref{fig:BunnyLevels}.
We observe that the level sets of the learned object are sufficiently accurate near the surface of the object.
The even spacing of level sets is important for collision detection, as the network gradient is used in the computation of the collision normal.

In the rest of this subsection, we first show in \autoref{sec:synthetic_bunny} a synthetic benchmark comparing the speed to compute the signed distance using our approach as well as with a standard BVH implementation from libigl \cite{libigl}.
Then we discuss in \autoref{sec:hypothetical_bunny} the performance of our current prototype and its hypothetical performance given the numbers from the synthetic benchmarks.

\subsubsection{Synthetic Benchmark}
\label{sec:synthetic_bunny}

With the synthetic benchmark, we compare the performance of our network evaluation against the BVH query implemented in libigl.
This benchmark consists of querying each method with points uniformly distributed within the bounding box of the model.
We include two network evaluation methods: CPU and GPU.
Since the network is trained on the GPU with PyTorch, it is trivial to evaluate on the GPU.
However, since the rest of the code for our application prototypes are written in C++ on the CPU, our current prototypes use CPU network evaluation code also written in C++.

\figBunnyLRQvsT{}

\autoref{fig:FRBunnyQvsT} shows the amount of time it takes to query the signed distance for the bunny in the haptics application, using libigl's BVH on the CPU, our network evaluation code on the CPU, and our network evaluation code on the GPU.
We assume that the bunny is deforming every time step, and so we require the BVH to first build the hierarchy and then make the query.
On the other hand, our approach simply evaluates the network.
(Overheads will be discussed next in \autoref{sec:hypothetical_bunny}.)

If the number of query points is small, then our network-based approach works well because it does not require an update.
However, as the number increases, BVH becomes faster, since the overhead of updating the hierarchy is amortized over a large number of queries.
Nevertheless, we find that the CPU evaluator is faster than BVH for up to about 300 query points, and the GPU evaluator is faster to about 10,000 query points, after which they stay neck-to-neck.
Looking at the slope of this log-log plot, we confirm that the asymptotic cost is linear (slope of 1) for all three methods.


\subsubsection{Current \& Hypothetical Performance}
\label{sec:hypothetical_bunny}

As mentioned earlier, our current prototype is written with C++ and runs on the CPU.
In this section, we combine our current collision performance numbers with the synthetic numbers to arrive at our hypothetical numbers.

In \autoref{sec:synthetic_bunny}, we showed that the neural evaluation of query points is faster than BVH when evaluating fewer than ${\sim}300$ points with CPU evaluation and ${\sim}10,000$ with GPU evaluation.
These synthetic performance numbers, however, omit the computation of the ID of the colliding triangle from the volumetric object's mesh corresponding to a given query point.
In our CPU implementation, we perform a linear scan over all triangles for this computation---we cast a ray from the query point along the network gradient (\ie collision normal) in order to determine which triangle is in collision with the query point.

If we perform this linear scan on the CPU after running the network evaluator on the GPU, we arrive at the worst case hypothetical performance numbers, shown in \autoref{tab:hypothetical_bunny}.
These numbers were obtained by computing the time it takes to perform the linear scan on the CPU and adding them to the GPU network evaluation time.
By our benchmarks, we find that moving the CPU network evaluation to the GPU, while keeping the linear scan on the CPU, would enable us to outperform a BVH when querying up to ${\sim}500$ to ${\sim}1,000$ points.

\begin{table}[]
\caption{Hypothetical performance numbers (in ms) for evaluating the signed distance for the bunny mesh on the GPU and then running the linear scan over the triangles on the CPU.}
\label{tab:hypothetical_bunny}
\begin{tabular}{r|cccccc} 
\textnormal{Num. Queries}  & 128 & 256 & 512 & 1024 & 2048 \\ \hline 
\textnormal{BVH Time}  & 4.88 & 6.46 & 7.25 & {\bf 10.55} & {\bf 17.18}\\  
\textnormal{Hyp. GPU Neural Time}  & {\bf 2.64} & {\bf 4.21} & {\bf 5.49} & 12.81 & 23.03
\end{tabular}
\end{table}

Next, we can hypothesize about the performance numbers when the linear scan is also performed on the GPU.
In calculating the numbers for the CPU linear scan in \autoref{tab:hypothetical_bunny}, we observed that this linear scan is highly parallelizable---we obtained near linear performance gains with 6 cores.
Therefore, we expect that moving this linear pass to the GPU would provide an even greater speedup, making our \textit{all-GPU} hypothetical numbers be closer to \autoref{fig:FRBunnyQvsT}.

\subsection{Cloth with Skinning}
\label{sec:results_cloth}

The character mesh has 27,999 vertices and 17,140 triangles; the cloth mesh has 1,404 vertices and 2,580 triangles.
We use 55 joint angles as the reduced code for the network.
For training, we use 400 poses with 10,000 SDF samples per pose.
The 400 poses were taken from two different animation files of the same character.
The visualizations of the levelsets of the learned SDF are shown in \autoref{fig:PersonLevels}.

\figPersonLevelsSmallNet{}

\figPersonQvsT{}

Similarly to our haptics application, a synthetic benchmark is run to compare the libigl BVH querying against our SDF evaluation on a learned character model.
As seen in \autoref{fig:PersonQvsT}, the advantage of our method over BVH is larger with the character than with the bunny.
This is because the character has many more triangles compared to the bunny, and so the initial overhead of building the hierarchy cannot be overcome as quickly.
We find that when evaluating up to around 3,000 points, our CPU method outperforms the BVH implementation, and when moving to the GPU, we gain a significant advantage and stay ahead.

We note that there are specialized BVH update methods for linear blend skinning \cite{kavan2005fast}, which would be much more competitive than rebuilding the hierarchy every time step.
However, our method is completely agnostic to the skinning method and thus can handle any skinning method, such as dual quaternion skinning or optimized center of rotation skinning \cite{kavan2007skinning,Le2016}.

Unlike our haptics application, the movement of the character is determined by an animation, not simulation. 
Therefore, the point queries performed in our cloth simulation do not need to identify which triangles of the learned mesh are in collision.
We have found that the overhead involved with performing a query (outside of network evaluation) is negligible when triangle selection is not performed.
Thus, assuming triangle selection is not required, the performance of a hypothetical GPU implementation of our full collision library corresponds directly to the evaluation times presented in \autoref{fig:PersonQvsT}.

\section{Conclusion}
\label{sec:conclusion}

We presented a neural network-based approach for collision detection of deformable objects.
We showed that the reduced degrees of freedom of the deformable object work effectively as the code for the implicit representation.
For an FEM object we used linear modes, and for a skinned character we used joint angles.
For the haptics application, linear modes worked well, even with only 128 reduced degrees of freedom.
We also showed that the material parameters chosen for runtime simulation can be significantly different from those used for training.

Through synthetic benchmarks, we showed that compared to a standard BVH implementation, we can compute the signed distance more efficiently for up to several hundred query points if evaluated on the CPU, and several thousand query points if evaluated on the GPU.
Our current prototype uses CPU network evaluation, and for the haptics application, we also perform a linear scan on the CPU to determine which triangle was involved in the collision.
However, we showed hypothetical performance numbers for the case if the network evaluation were to be performed on the GPU, as well as if both the evaluation and the linear scan were to be performed on the GPU.

\subsection{Future Work}

We believe that there are many novel and exciting directions in which our current work can be extended.
An immediate goal of ours is to integrate a fully GPU-backed pipeline for collision detection, including the linear scan.
We also plan to eventually port the whole simulation pipeline to the GPU, including assembly and linear solve.
We also hope to further explore the limits of the reduced coordinates for collision, with regard to both the network hyperparameters and the code space.

For FEM, we plan to explore other reduced codes, such as nonlinear modes.
For cloth, we plan to utilize the interpolating capability of neural networks into other dimensions---for example, we could use a single network to interpolate between multiple characters with the same skeletal structure.

Finally, to make our neural approach more scalable, we also plan to investigate using a hierarchical approach.
We expect that just a few levels would be highly beneficial to our neural collision framework.



\bibliographystyle{ACM-Reference-Format}
\bibliography{neural-collision.bib}

\end{document}